# super-reflective data:

## speculative imaginings of a world where data works for people

Max Van Kleek
University of Oxford
max.van.kleek@cs.ox.ac.uk

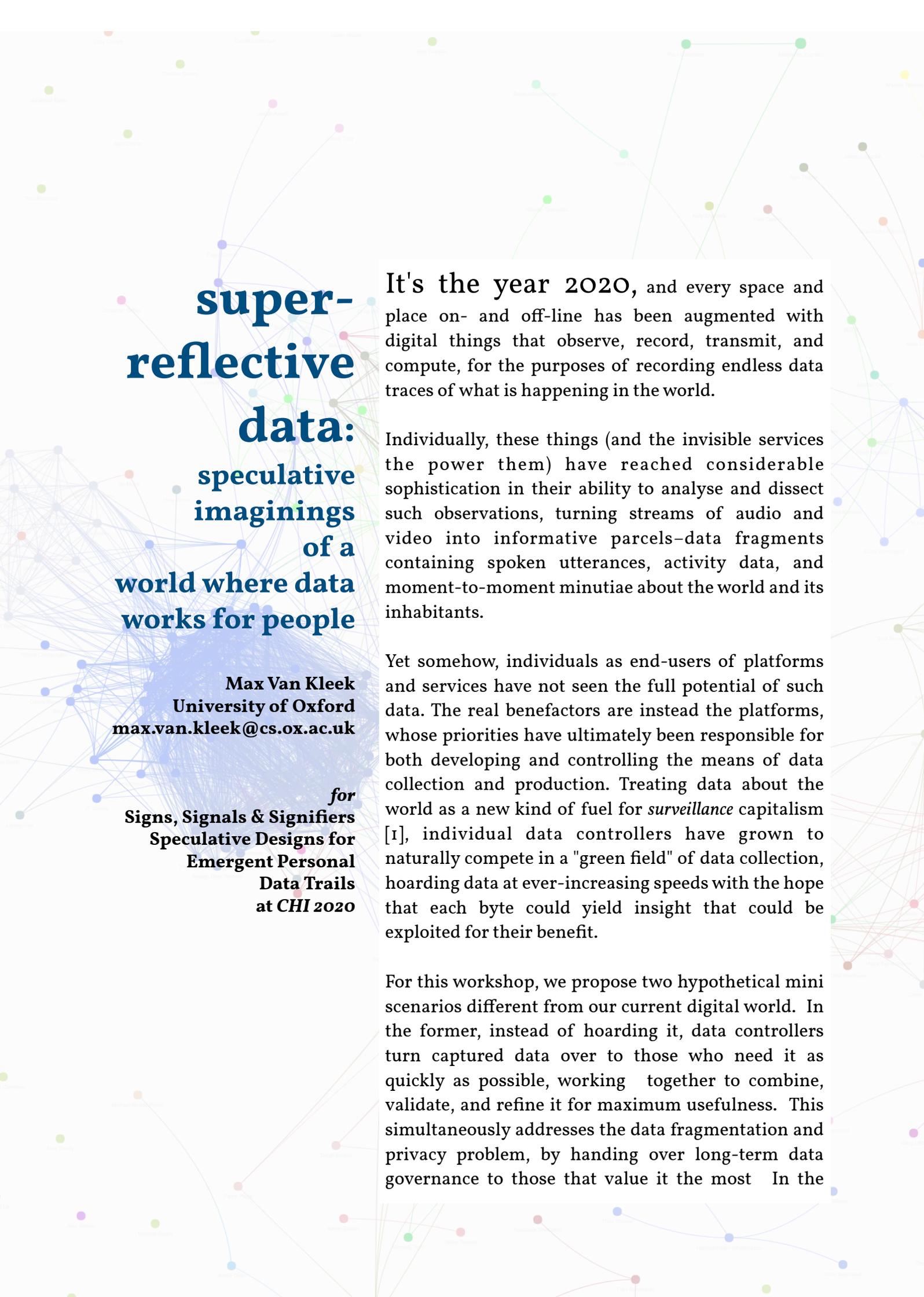

**It's the year 2020,** and every space and place on- and off-line has been augmented with digital things that observe, record, transmit, and compute, for the purposes of recording endless data traces of what is happening in the world.

Individually, these things (and the invisible services the power them) have reached considerable sophistication in their ability to analyse and dissect such observations, turning streams of audio and video into informative parcels–data fragments containing spoken utterances, activity data, and moment-to-moment minutiae about the world and its inhabitants.

Yet somehow, individuals as end-users of platforms and services have not seen the full potential of such data. The real benefactors are instead the platforms, whose priorities have ultimately been responsible for both developing and controlling the means of data collection and production. Treating data about the world as a new kind of fuel for *surveillance* capitalism [1], individual data controllers have grown to naturally compete in a "green field" of data collection, hoarding data at ever-increasing speeds with the hope that each byte could yield insight that could be exploited for their benefit.

For this workshop, we propose two hypothetical mini scenarios different from our current digital world. In the former, instead of hoarding it, data controllers turn captured data over to those who need it as quickly as possible, working together to combine, validate, and refine it for maximum usefulness. This simultaneously addresses the data fragmentation and privacy problem, by handing over long-term data governance to those that value it the most In the latter, we discuss ethical dilemmas using the long-term use of such rich data and its tendency to cause people to relentlessly optimise.

**Scene 1: Data Reflections in Millions of Mirrors –**
When citizens of a city in 2030 move about town, they rarely have to worry about ever remembering anything anymore-even charging or wearing their wearables or smartphone. Someone (or, more likely *something*) else will do it for them. Should there be a need to recollect an event, such as if you were taken ill or became injured, something will have almost certainly witnessed it.

Without even being asked, the capturer would have processed, analysed, and offered up this data to your own personal data infrastructure for your use. This is because all sensing infrastructure, public or private are now bound to the national and international Ethical Data Codes of Conduct drawn together to not only protect people from misuses of data, but to make sure to make potentially valuable data about subjects to them immediately.

Since these Ethical Data codes require data retention acts to be justified under the basis of some direct benefit to individuals, most infrastructure controllers relinquish captured data by handing it over to those that can benefit the most from having it. This is beneficial for many reasons; not the least of which that while 99.99% of captured fairly useless, that final fractional percent can be tremendously valuable, but often only to very certain individuals - in the right contexts and with the right needs.

This led to the establishment of a data access forum by which the availability of data is registered, so that those who could benefit could easily find and request it ~ demonstrating, of course, that they have a valid need and that such access wouldn't violate others' privacy. Fortunately, with the advances of personal data demultiplexing, all privacy implications of data can be "taken apart" and factored into components that let enable benefits without privacy risks.

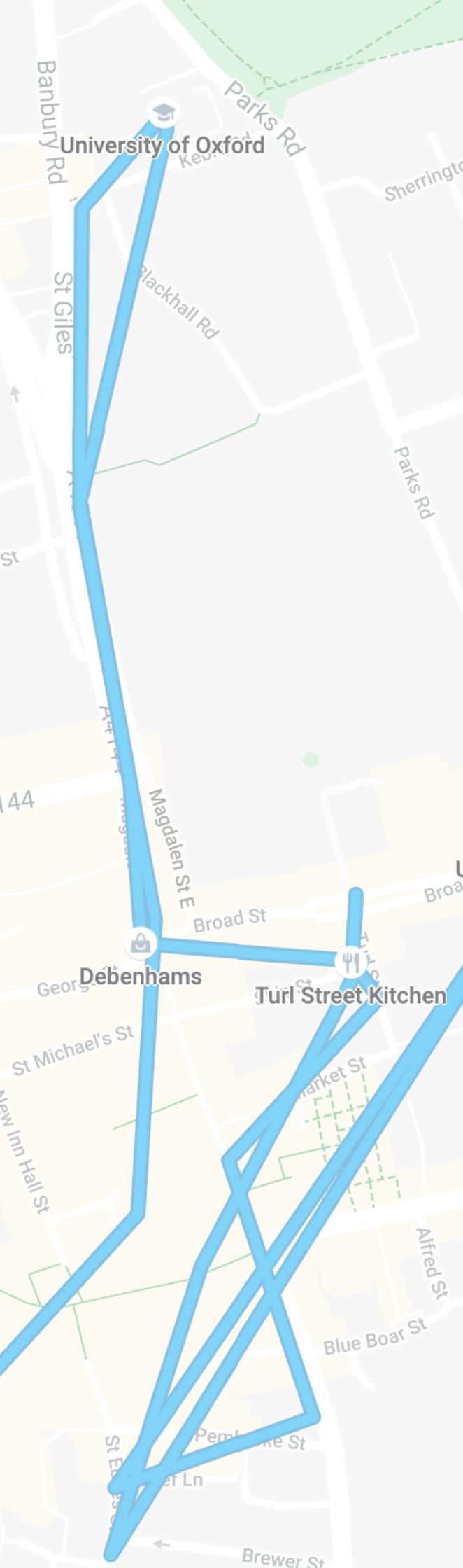

This new ethical data order has already seen profound effects on digital society. First of all, end-users who previously had little or no access to their own data, or who, at best, had small useless fragments, are now awash in unified, enriched data about themselves, being volunteered by infrastructure and platform providers. This gives end-users the benefit of allowing them to trivially piece together rich multifaceted diaries of their lives- social, physical, and even intellectual/psychological, unlocking significant possibilities for new classes of *life mining* applications that find opportunities for people to reflect upon and optimise their lives with minimal cost and effort. The economic value generated by such applications has demonstrated such significant returns on infrastructure investment that it has driven a burgeoning infrastructure market of providers rushing to expand coverage and add ever-more sophisticated sensing apparatuses. Large-scale data breaches have become exceedingly rare, because casual data controllers typically pass data over to subjects rather than be responsible for it themselves. Finally, and perhaps most happily, privacy violations have become much rarer thanks to platforms having to justify that their use of data promotes the well-being of data subjects, and the overwhelming transition of the personal data governance to end-users.

**Scene 2: Data Trails of Highly Successful People** – The availability of data about people's lives has now made it possible for people to live optimally, by cutting the guesswork out of life. Just as the early vanguard of Quantified Self athletes found that they could eke a bit of extra performance out of following the tried-and-true self-experiments, now life traces make it possible for people to live optimally in ways beyond physical aptitude simply by following in the literal footsteps of their data forebearers.

Thanks to the millions of life traces submitted by the world's fully connected population, we can now quantify how optimally a person is living their life. Want to be 2% fitter, happier, or more successful?

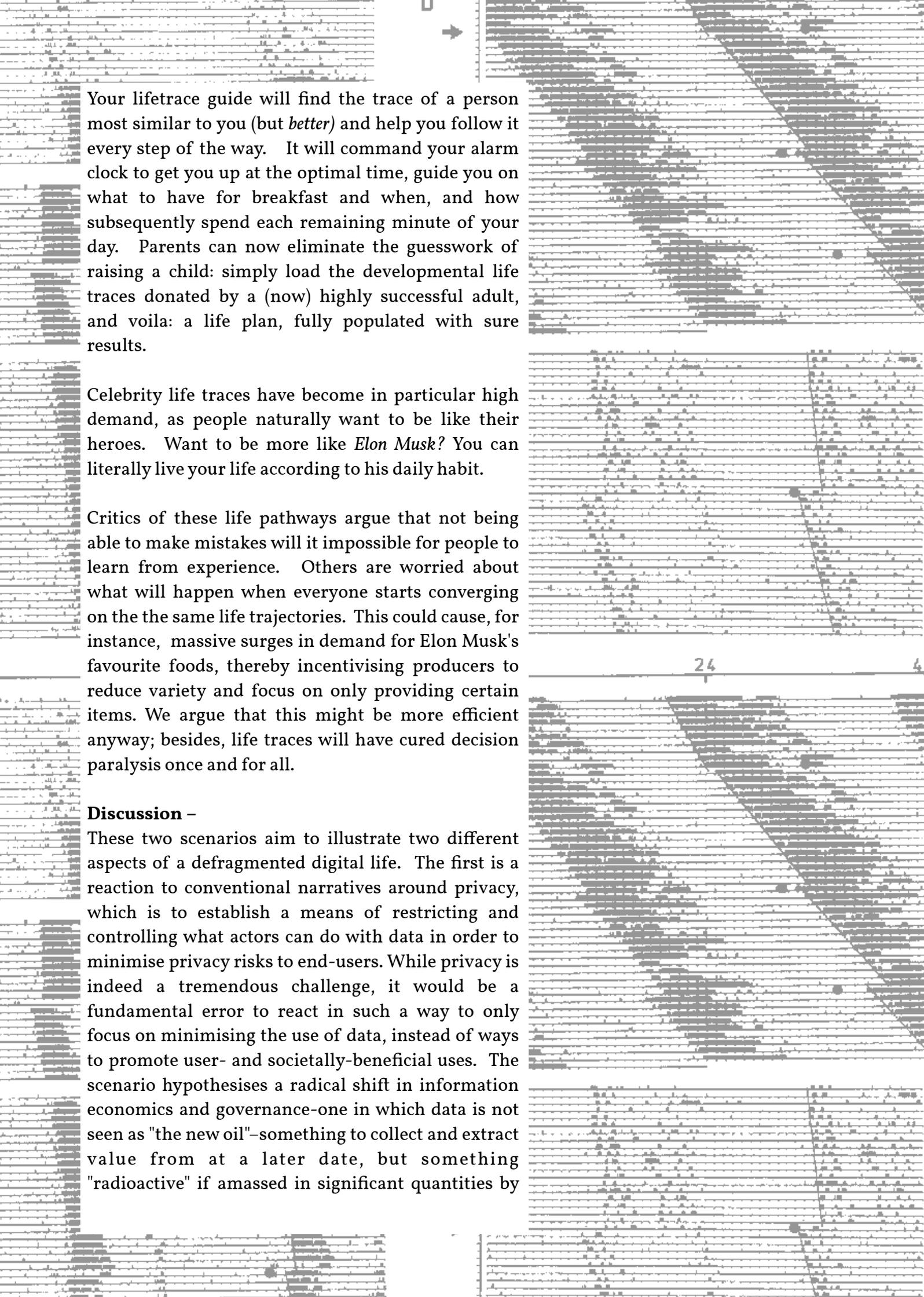

Your lifetrace guide will find the trace of a person most similar to you (but *better*) and help you follow it every step of the way. It will command your alarm clock to get you up at the optimal time, guide you on what to have for breakfast and when, and how subsequently spend each remaining minute of your day. Parents can now eliminate the guesswork of raising a child: simply load the developmental life traces donated by a (now) highly successful adult, and voila: a life plan, fully populated with sure results.

Celebrity life traces have become in particular high demand, as people naturally want to be like their heroes. Want to be more like *Elon Musk?* You can literally live your life according to his daily habit.

Critics of these life pathways argue that not being able to make mistakes will it impossible for people to learn from experience. Others are worried about what will happen when everyone starts converging on the the same life trajectories. This could cause, for instance, massive surges in demand for Elon Musk's favourite foods, thereby incentivising producers to reduce variety and focus on only providing certain items. We argue that this might be more efficient anyway; besides, life traces will have cured decision paralysis once and for all.

**Discussion** –
These two scenarios aim to illustrate two different aspects of a defragmented digital life. The first is a reaction to conventional narratives around privacy, which is to establish a means of restricting and controlling what actors can do with data in order to minimise privacy risks to end-users. While privacy is indeed a tremendous challenge, it would be a fundamental error to react in such a way to only focus on minimising the use of data, instead of ways to promote user- and societally-beneficial uses. The scenario hypothesises a radical shift in information economics and governance-one in which data is not seen as "the new oil"–something to collect and extract value from at a later date, but something "radioactive" if amassed in significant quantities by intermediaries, and safe if passed to, and governed by data subjects themselves.

One of the main benefits of such an approach will be data fragments from multiple sources can be unified, and brought together within a context which the most meaningful: that of the data subject. For instance, the data subject is more likely to be able to understand and know "why" they were doing something at a certain time and place better than anyone else - and use this to discern whether the data are significant on this basis.

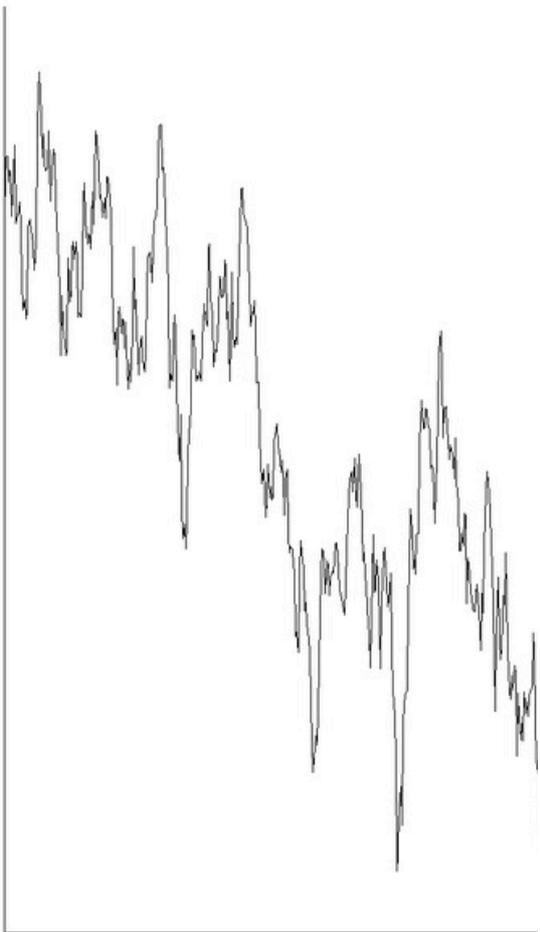

The second illustrates a more fundamental challenge with an increasingly data-driven society: the relentless seduction of data as a means of optimisation and measuring "progress". We have already started to see the effects of relentless quantification and data-trace based competition within fitness and well-being apps, which establish harmful normative notions of "wellness" [2] and get drive people towards unprecedented convergent fitness-related addictions and obsessions [3,4]. We will need more radical thinking to understand how to offset our tendency to over-quantify and measure life outcomes, perhaps explicitly by exploring the value of "non-optimal" living: curiousity, variety, exploration and the value of the journey rather than the result.

## References –